\documentclass[5p,times,twocolumn]{elsarticle}

\usepackage{amsmath,amssymb}
\biboptions{comma,sort&compress}
\usepackage[utf8]{inputenc}
\usepackage{graphicx}
\usepackage{bm}
\usepackage{hyperref}
\hypersetup{
colorlinks = true,
linkcolor = blue,
anchorcolor = blue,
citecolor = blue,
filecolor = blue,
urlcolor = blue
}
\usepackage[dvipsnames]{xcolor}
\usepackage{algorithmicx}
\usepackage[ruled]{algorithm}
\usepackage[noend]{algpseudocode}
\usepackage{enumitem}
\usepackage{listings}
\usepackage{xcolor}
\usepackage{tikz}
\usetikzlibrary{shapes}
\usetikzlibrary{positioning,arrows}
\usepackage{scalefnt}

\graphicspath{{figs/}}

\newcommand{\grid}{{GRID}}
\newcommand{\tensorcalc}{{TensorCalc}}

\journal{Computer Physics Communications}
\begin{document}
\begin{frontmatter}

\title{Efficient calculation of thermodynamic properties of baryon-rich QCD matter\\from heavy-ion transport models}

\author[first,second]{Lipei Du}
\ead{ldu2@lbl.gov}
\affiliation[first]{organization={Department of Physics, University of California},
            city={Berkeley},
            postcode={94270}, 
            state={CA},
            country={USA}}
\affiliation[second]{organization={Nuclear Science Division, Lawrence Berkeley National Laboratory},
            city={Berkeley},
            postcode={94270}, 
            state={CA},
            country={USA}}
\date{\today}
\begin{abstract}
This study presents the MATRICS framework (Modeling Aggregated Tensors for Relativistic Ion Collision Simulations) that implements modular workflows to enable parallel execution of particle generation, grid construction, and tensor calculations for heavy-ion collisions. It introduces an efficient approach to calculating the space-time distribution of the energy-momentum tensor and charge currents from discrete particles generated by transport models. By dynamically adjusting grid resolution based on particle density and clustering particles into representative super-particles, MATRICS optimizes computational efficiency while maintaining high physical accuracy. The framework can also provide a thermodynamic background for electromagnetic thermal emission calculations or serve as initial conditions for hydrodynamic evolution. It offers a powerful tool for exploring the thermodynamic properties of QCD matter at high baryon densities, making it well-suited for large-scale simulations in heavy-ion collision studies.
\end{abstract}


\end{frontmatter}


\section{Introduction}

Strongly interacting matter exhibits complex phases, and mapping the structure of the Quantum Chromodynamics (QCD) phase diagram is a central goal of the nuclear physics community \cite{Braun-Munzinger:2008szb,Sorensen:2023zkk}. Theoretical predictions suggest the existence of a critical point at nonzero chemical potential, but confirming its existence and precise location remains a significant challenge \cite{Stephanov:1998dy,Stephanov:1999zu}. Relativistic heavy-ion collisions at various beam energies provide the primary experimental method to probe the QCD phase diagram, with lower beam energies effectively exploring regions of higher baryon chemical potential \cite{Bzdak:2019pkr,Du:2024wjm,An:2021wof}.

The Beam Energy Scan (BES) program at RHIC has been instrumental in exploring the QCD phase diagram and searching for the critical point, extending center-of-mass energies down to $\sqrt{s_{\text{NN}}}=3$ GeV \cite{Bzdak:2019pkr,Du:2024wjm}. Data analysis is ongoing, with high-order cumulants of protons expected to serve as sensitive probes of critical phenomena \cite{Stephanov:2008qz, Stephanov:2011pb}. Current efforts focus on collisions below $\sqrt{s_{\text{NN}}}=20$ GeV, where model-to-data comparisons are crucial for substantiating the existence of the critical point, given the highly dynamic and inhomogeneous nature of QCD matter produced in these collisions. Another key challenge is determining the beam energy threshold below which nuclear collisions no longer produce a deconfined system, transitioning to a hadronic gas-dominated regime \cite{STAR:2021yiu,Du:2024wjm}.

Multi-stage hydrodynamic models have successfully described many experimental observations at high center-of-mass energies ($\sqrt{s_{\text{NN}}} \gtrsim 200$ GeV) \cite{Bernhard:2019bmu,JETSCAPE:2020shq,Nijs:2020ors}. However, at lower beam energies ($\sqrt{s_{\text{NN}}} \lesssim O(10)$ GeV), the prolonged initial stage of nuclear penetration complicates modeling \cite{Bzdak:2019pkr,Du:2024wjm,An:2021wof}. Two major uncertainties arise: whether the system achieves sufficient thermalization for hydrodynamic descriptions to apply, and how to construct the sophisticated space-time dependent initial conditions required for hydrodynamic evolution  \cite{Shen:2017bsr, Du:2018mpf, Shen:2017ruz,Du:2019obx}. Dynamically modeling energy loss and charge stopping can help address these challenges \cite{Shen:2017bsr,Shen:2022oyg}, as can the use of transport models to simulate initial-stage hadronic scatterings and construct continuous energy-momentum tensors and charge currents for hydrodynamic initialization \cite{Petersen:2008dd,Karpenko:2015xea}. Transport models are also widely used to simulate low-energy collisions from start to finish \cite{Bleicher:2022kcu,TMEP:2022xjg}.

Regardless of the approach, whether constructing hydrodynamic initial conditions or investigating thermodynamic properties within transport models, deriving continuous space-time distributions from discrete particles is essential \cite{Oliinychenko:2015lva,Inghirami:2022afu}. This typically involves applying smearing kernels to particles and aggregating their contributions across space-time, a process that is computationally demanding, especially when repeated for thousands of events.

The MATRICS framework (Modeling Aggregated Tensors for Relativistic Ion Collision Simulations) addresses these challenges by introducing innovative computational techniques to optimize the calculation of energy-momentum tensors and charge currents from discrete particles. Key contributions include dynamic particle clustering methods, such as box clustering, which group particles into representative super-particles, significantly reducing computational costs while preserving physical accuracy when combined with smearing kernels. The framework also integrates parallel processing and modular workflow design, enabling efficient handling of large-scale simulations with thousands of collision events. By combining these advancements, MATRICS achieves a balance between computational efficiency and physical accuracy, making it a powerful tool for studying the space-time evolution of QCD matter thermodynamic properties in heavy-ion collisions, particularly in high baryon density regions.

\tikzstyle{decision} = [diamond, draw, fill=yellow!20, 
    text width=7em, text badly centered, node distance=3cm, inner sep=0pt]
\tikzstyle{block} = [rectangle, draw, fill=blue!20, 
    text width=20em, text centered, rounded corners, minimum height=8em]
\tikzstyle{cloud} = [draw, ellipse,fill=green!20, node distance=3cm,
    minimum height=2em]
\tikzstyle{line} = [draw, very thick, color=black, -latex']

\begin{figure*}[!htbp]
\begin{center}
\resizebox{0.95\textwidth}{!}{
\begin{tikzpicture}[node distance = 10cm, every node/.style={font=\large}]
    \node [block] (setup) {Set up and run SMASH \\ in parallel across folders\\(SMASH)};
    \node [block, right of=setup] (globalGrid) {Determine global regular or adaptive grid \\ using representative events\\(\grid{})};
    \node [block, right of=globalGrid] (calcTensor) {Calculate and average \\ energy-momentum tensors and charge currents in parallel\\(\tensorcalc{})};

    \path [line] (setup) -- (globalGrid);
    \path [line] (globalGrid) -- (calcTensor);

\end{tikzpicture}
}
\end{center}
\caption{Flow chart of the MATRICS framework, including parallel SMASH execution, global grid determination, and energy-momentum tensor calculation and averaging.\label{fig:WorkflowChart}} 
\end{figure*}
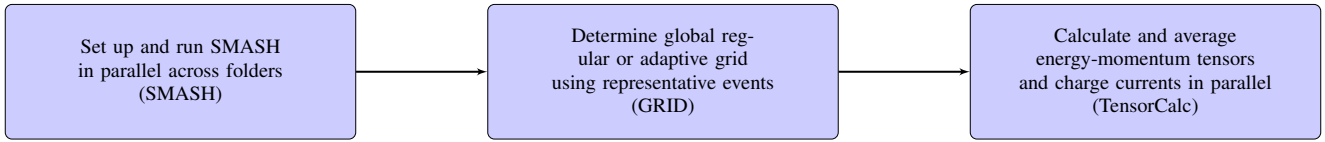

\section{Workflow design and implementation}

This section outlines the structure and modularization of the workflow, along with the high-level scripts utilized for setting up and executing the workflow.

\subsection{Code structure and modularization}

The workflow for MATRICS, illustrated in Fig.~\ref{fig:WorkflowChart}, is designed to optimize the simulation of particle production from heavy-ion collisions (SMASH), the assessment of regular or adaptive space-time grids (\grid{}), and the calculation of the space-time evolution of continuous tensors and currents (\tensorcalc{}). This structured approach ensures modularity, facilitates parallel execution, and promotes efficient data handling, which is particularly crucial given the substantial volume of data generated by these simulations.

SMASH, a transport model event generator, simulates particle production and rescattering in heavy-ion collisions \cite{SMASH:2016zqf,wergieluk_2024_10707746}. As it is not inherently parallelized, the workflow operates multiple independent instances of SMASH simultaneously across various directories to generate events in parallel. Typically, these SMASH events are configured within the same centrality class, although they may vary in impact parameters from event to event. This parallel execution is accomplished through multiple threads, with each thread launching a separate SMASH instance. Such parallelism significantly reduces overall computation time, enabling the efficient generation of thousands of collision events.

Following the simulation of events and the generation of particle lists, the next stage constructs space-time grids for calculating the space-time evolution of the energy-momentum tensor. A dedicated module named \grid{} is introduced, which isolates the grid determination process from the main tensor calculation. This step is essential because the energy-momentum tensors from different events often require averaging, necessitating a global space-time grid applicable across all events. To address this, the particle lists from all SMASH runs are aggregated and utilized by \grid{} to establish a four-dimensional space-time grid. This centralized approach ensures global consistency, capturing the overall space-time structure of all collision events. The grid calculation is designed to be flexible, allowing for the construction of both regular and adaptive grids. Further details on grid determination will be discussed in Sec.~\ref{sec:grid}.

The final stage of the workflow involves calculating the energy-momentum tensors and charge currents from the particle lists using \tensorcalc{} which is the framework's core. Similar to SMASH, \tensorcalc{} operates in parallel, with each instance functioning within its designated folder. A \tensorcalc{} instance reads the particle lists generated by SMASH in the same folder, along with the global space-time grid information, and then iteratively calculates the tensors over the grid. To expedite computation, events can be processed in parallel, with each thread calculating the energy-momentum tensor for a subset of events and updating the global tensor in a thread-safe manner. Once all events are processed, the average energy-momentum tensors are computed across all events or within subsets of events. \tensorcalc{} employs advanced algorithms, including particle clustering and the use of representative particles, to substantially reduce computational costs while maintaining high physical accuracy (Sec.~\ref{sec:clustering}).

\subsection{High-level workflow scripts}

The overall workflow for MATRICS, as described in the previous section, is implemented through two high-level Python scripts, \texttt{matrics\_setup\_control.py} and \texttt{matrics\_workflow\_driver.py}.\footnote{
These scripts are inspired by the iEBE-MUSIC framework \cite{iebe}.
}
These scripts manage the comprehensive workflow and ensure the seamless integration of various modules. They are essential for setting up, executing, and managing the project's computational components, including SMASH, \grid{}, and \tensorcalc{}, in an efficient and automated manner.

The \texttt{matrics\_setup\_control.py} script is tasked with establishing the project’s working directory and maintaining a well-organized structure. It creates a folder hierarchy with subdirectories to facilitate the parallel execution of multiple SMASH and \tensorcalc{} instances, based on the specified number of threads, along with a dedicated directory for a single instance of \grid{}. This script also generates and links configuration files for each module, utilizing user-defined parameters from \texttt{matrics\_config\_user.py} alongside default values from \texttt{matrics\_config\_default.py}. By systematically organizing and linking code and parameter files within each subfolder, the script ensures that each instance of SMASH, \grid{}, and \tensorcalc{} operates with consistent and appropriate settings. Furthermore, \texttt{matrics\_setup\_control.py} produces job submission scripts to streamline the workflow's execution on high-performance computing clusters.

The \texttt{matrics\_workflow\_driver.py} script coordinates the execution of the computational modules in the designated sequence. It initiates the parallel execution of SMASH simulations across the subfolders, thereby efficiently generating particle lists. Once the SMASH outputs are available, the script aggregates these particle lists for the \grid{} module, which determines a global adaptive space-time grid based on representative event samples. This grid is essential for the subsequent calculations of the energy-momentum tensor. The \grid{} code is executed once to generate this global grid, which is subsequently utilized to calculate the spatial and temporal smearing of particles in the \tensorcalc{} instances. The parallel execution of \tensorcalc{} instances computes energy-momentum tensors using the pre-established grid from \grid{}, followed by averaging these tensors across all events to yield the final results.

Both high-level scripts are designed with flexibility and efficiency at their core. They allow for dynamic parameter adjustments through user-specified configurations in \texttt{matrics\_config\_user.py}, enabling users to tailor the simulations for various scenarios. The scripts also implement mechanisms to verify the existence of intermediate outputs, ensuring that computations progress smoothly and efficiently, thereby avoiding redundant or unnecessary re-execution of completed tasks. By automating the workflow setup and execution, these scripts significantly reduce the complexity associated with managing large-scale heavy-ion collision simulations.

\section{Algorithms and optimizations}\label{sec:algorithm}

This section describes two major methods, grid calculation and particle clustering, which are essential for reducing computational costs while maintaining physical accuracy.

\subsection{Grid calculation}\label{sec:grid}

The design of the grid is a critical aspect of this workflow, as it dictates how the space-time distribution of the energy-momentum tensor is calculated. Computationally, it significantly impacts the overall cost, as it constitutes one of the major loops in the tensor calculation. Physically, it also affects the accuracy of thermodynamic properties and their gradients. The workflow supports two types of grids within \grid{}: regular and adaptive. Each grid type has specific design considerations and advantages based on the simulation requirements.

\subsubsection{Regular grid}\label{sec:regular}
The regular grid represents a straightforward implementation, maintaining a fixed resolution throughout the entire evolution, characterized by uniform grid sizes and a constant number of cells in each dimension (spatial and temporal). The primary advantage of the regular grid lies in its predictability and ease of use, while it avoids potential issues with discontinuities or irregularities in grid spacing that may arise with adaptive grids. This grid type is particularly useful for benchmarking purposes or when outputs must be compatible with other simulation codes that require a uniform grid format. For instance, it is employed when constructing initial conditions for hydrodynamic evolution from discrete particles of a transport model \cite{Du:2018mpf,Du:2019obx}.

In heavy-ion collisions, the systems expand significantly over time, necessitating a grid that encompasses all particles of interest, particularly at late times when the volume is large \cite{Oliinychenko:2015lva}. Within \grid{}, the total grid size in each dimension is determined by the largest enclosing box capable of containing all relevant particles. \tensorcalc{} also provides users with the option to define a regular grid, enabling it to operate as a standalone code when necessary, without relying on \grid{}.

However, regular grids may lead to inefficiencies for two reasons in the context of heavy-ion collisions. First, they assign the same resolution to low-density regions (late stages) as they do to high-density regions (early stages). Second, they allocate the same total grid size to compact states (early stages) as they do to expanded states (late stages).\footnote{
This issue is particularly significant at low collision energies, where Milne coordinates are less preferred compared to Cartesian coordinates. When Cartesian coordinates are used, the grid requires a substantially larger size in the $z$-direction to enclose the expanded states, which corresponds to the beam direction of the collisions \cite{Oliinychenko:2015lva}. In contrast, with Milne coordinates, the grid size in the beam direction is limited by the beam rapidity.
}  
Both of these factors can result in excessive computational costs in areas that may be sparse or even empty of particles.

\subsubsection{Adaptive grid}\label{sec:adaptive}
To address the limitations of a regular grid, the adaptive grid is introduced as a more advanced approach. This grid dynamically adjusts its resolution based on the spatial and temporal distribution of particle densities, as well as its overall size to accommodate the expanding system. A key feature of the adaptive grid is its ability to provide higher resolution in regions of high particle density (such as the early stages of collisions) while maintaining lower resolution in sparse regions (such as the late stages). Furthermore, it allocates smaller grid sizes when the system is compact (early stages) and larger grid sizes as the system expands (late stages), ensuring that the grid remains optimally sized for the evolving state of the collision.

The adaptive grid is constructed by analyzing particle densities using their space-time coordinates, determining cell sizes that are inversely proportional to local particle density. This allows for dynamic adjustments of the grid structure at different time steps, accounting for both system expansion and variations in density over time. As a result, the adaptive grid effectively captures the complex, evolving structures of heavy-ion collisions---such as high-density or highly inhomogeneous regions---while minimizing unnecessary computational effort in smoother or less dense areas. By aligning the grid with the underlying particle distribution, the adaptive grid ensures that computational resources are focused on regions of significant physical activity, reducing costs while maintaining an accurate representation of the collision dynamics.

The adaptive grid is particularly valuable when the constructed energy-momentum tensor is not utilized as initial conditions for hydrodynamic evolution. For instance, it is advantageous when estimating the hydrodynamization or thermalization of a system by analyzing the isotropization of the energy-momentum tensor \cite{Oliinychenko:2014tqa,Inghirami:2019muf}, or when calculating electromagnetic emissions from a macroscopic perspective \cite{Endres:2015fna,Savchuk:2022aev,Du:2024pbd}. The adaptive grid is also ideal for detailed investigations of localized phenomena in heavy-ion collisions, where local properties such as gradients are critical \cite{Du:2023efk,Du:2023gnv,Li:2023kja}. In these scenarios, a regular grid is unnecessary, allowing the adaptive grid to reduce computational expenses significantly.

The dual design of adaptive and regular grids enhances the workflow's flexibility, allowing it to be tailored to the specific needs of different physics applications. In both grid types, the designs aim to balance computational resources with the accuracy of tensor calculations. By accommodating both grid types, the workflow can address a wide range of simulations, from high-precision studies of small-scale physics to broader analyses of macroscopic properties.

\subsection{Particle clustering}\label{sec:clustering}

Handling large numbers of particles across multiple events at each space-time location constitutes a major computational bottleneck in tensor calculations. To reduce this complexity, we implement methods that group particles into clusters and treat these clusters as single effective super-particles in \tensorcalc{}. This approach can be interpreted as a form of coarse-graining, which is physically justified in the context of heavy-ion collisions. Particles produced within specific regions of space-time often exhibit correlations due to the underlying collective hydrodynamic behavior. Consequently, particles occupying spatially close positions typically possess similar velocities and momenta, allowing their combined contribution to the energy-momentum tensor to be reasonably approximated by a single effective particle. The effectiveness of using representative particles to preserve the underlying physics will be discussed in Sec.~\ref{sec:physics_accuracy}.

When focusing on large-scale features rather than fine-grained structures, grouping particles within small space-time regions does not significantly compromise the overall accuracy of the physics being represented. This approach is particularly advantageous for investigating quantities such as temperature and baryon chemical potential, which inherently smooth out small-scale fluctuations. It is especially useful when deriving event-averaged thermodynamic properties from the event-averaged energy-momentum tensors and charge currents. 

We explore two types of clustering methods: the first, box clustering, which clusters particles into space-time boxes (detailed in Sec.~\ref{sec:boxclustering}), and the second, dynamical clustering, which groups particles more dynamically into energy-weighted centroids (discussed in Sec.~\ref{sec:dynclustering}).

\subsubsection{Box clustering}\label{sec:boxclustering}

Box clustering involves grouping particles into space-time boxes, which can be implemented in two distinct ways: box clustering on a predefined regular grid (``regular box cluster'') and box clustering on a dynamical grid  (``dynamical box cluster''). They differ significantly in their implementation and resulting particle distributions. 

The first method, ``regular box cluster'', employs a fixed grid with uniform spacing defined by parameters $\text{d}t$, $\text{d}x$, $\text{d}y$, and $\text{d}z$.\footnote{
This is typically chosen to match the uniform space-time grid discussed in Sec.~\ref{sec:regular}, which is used for calculating the energy-momentum tensor and charge currents.
} 
Particles are assigned to grid cells based on predefined boundaries, ensuring a uniform spatial organization. It requires iterating over both particles and predefined grid points, leading to a computational cost of $O(N_{\text{particles}} \times N_{\text{grid}})$. This is because it must check each particle against every grid point, even if many grid points are unoccupied. Additionally, this approach can lead to artifacts in regions with low particle density, as empty grid cells are still included in the clustering process. The resulting clusters are rigidly aligned with the grid, which may not accurately reflect the true particle distribution, especially in sparse regions. 

\alglanguage{pseudocode}
\renewcommand{\algorithmicrequire}{\textbf{Input:}}
\renewcommand{\algorithmicensure}{\textbf{Output:}}
\begin{algorithm}[t!]
\small
\caption{Space-time box clustering with dynamical grid}
\label{Algorithm:SpaceTimeBoxClustering}
\begin{algorithmic}[1]
\Require \texttt{particles} - list of particles
\Require \hspace{.75cm} \texttt{dt, dx, dy, dz} - grid spacing in time and space
\Ensure \texttt{representativeParticles} - map of grid points to representative particles
\Function{$\mathbf{SpaceTimeBoxClustering}$}{\texttt{particles, dt, dx, dy, dz}}
    \State \texttt{boxMap} $\leftarrow$ empty hash map \Comment{Map grid points to particle lists}
    \State \texttt{representativeParticles} $\leftarrow$ empty hash map \Comment{Map grid points to representative particles}
    \For {\texttt{each particle in particles}}
        \State \texttt{gridPoint} $\leftarrow$ \Call{ComputeIndices}{\texttt{particle, dt, dx, dy, dz}} \Comment{Compute grid indices}
        \State \texttt{boxMap[gridPoint].push\_back(particle)} \Comment{Group particles into boxes}
    \EndFor
    \For {\texttt{each (gridPoint, boxParticles) in boxMap}}
        \State \texttt{representativeParticles[gridPoint]} $\leftarrow$ \Call{ComputeRepresentative}{\texttt{boxParticles}} \Comment{Compute representative particle}
    \EndFor
    \State \Return \texttt{representativeParticles}
\EndFunction
\Statex
\end{algorithmic}
  \vspace{-0.4cm}%
\end{algorithm}

In contrast, ``dynamical box cluster'' dynamically defines the grid based on the positions of the particles themselves.\footnote{
Note that this refers to the space-time grid of particle coordinates, which is different from the adaptive grid used for tensor calculations, discussed in Sec.~\ref{sec:adaptive}.
}
This method generates grid cells only where particles are present, creating a more adaptive and event-specific grid structure. The clustering process directly reflects the particle distribution, with dense regions forming more populated clusters and sparse regions potentially resulting in fewer or no clusters. This adaptability reduces the risk of artifacts from unoccupied grid cells and ensures that the clustering aligns more naturally with the actual particle density. As a result, the energy distribution in the ``dynamical box cluster'' better captures the true gradients in particle density, particularly in high-density regions. 

The efficiency of ``dynamical box cluster'' is achieved through the use of hash maps (\texttt{std::unordered\_map}) \cite{Mehlhorn2008}, which enable fast insertion and lookup operations with average $O(1)$ complexity per particle, leading to an overall complexity of $O(N_{\text{particles}})$. As outlined in Algorithm~\ref{Algorithm:SpaceTimeBoxClustering}, each particle is assigned to a grid cell by computing its space-time indices, and the hash map stores only non-empty grid cells and their associated particles. This approach avoids iterating over predefined grid points, significantly reducing computational overhead, especially for sparse particle distributions. The custom hash function ensures efficient grouping of particles into dynamically defined grid cells, while the computation of representative particles (see Sec.~\ref{sec:repparticles}) for each cell further enhances the method's adaptability. By processing only occupied cells, the clustering process scales linearly with the number of particles, making it highly efficient for large systems.

In summary, the ``regular box cluster'' method requires a computational cost of $O(N_{\text{particles}} \times N_{\text{grid}})$, while the ``dynamical box cluster'' scales as $O(N_{\text{particles}})$. This difference in scaling becomes particularly significant for large systems or sparse particle distributions, where the ``dynamical box cluster'' avoids unnecessary computations on empty grid cells. A detailed analysis of the computational efficiency of both methods, including their impact on the smearing step, is provided in Sec.~\ref{sec:computational_efficiency}. The ``dynamical box cluster'' not only improves computational performance but also ensures that the clustering process is more physically meaningful, as it directly reflects the underlying particle distribution, as will be demonstrated in Sec.~\ref{sec:physics_accuracy}. This makes it particularly suitable for applications where capturing fine-grained density gradients and reducing computational overhead are critical.

\subsubsection{Dynamical particle clustering}\label{sec:dynclustering}

We also explore a clustering approach that groups particles dynamically into energy-weighted centroids. Specifically, for each time step, we utilize the $k$-means clustering algorithm \cite{likas2003global} to partition the set of particles into distinct groups based on their spatial and momentum distributions. The $k$-means clustering algorithm is a widely used unsupervised machine learning technique that segments a dataset into $k$ distinct, non-overlapping clusters. The following outlines the step-by-step process of the algorithm (see Algorithm~\ref{Algorithm:SimpleKMeansClustering}):

\begin{description}[leftmargin=0.0cm, font=\normalfont, itemsep=-0.1em]
\item[Initialization of centroids:] The algorithm begins by randomly selecting $k$ particles from the dataset to serve as the initial centroids. These centroids represent the starting positions of the clusters. Conceptually, this step can be viewed as making an initial estimation of the ``center'' of various groups of particles in a spatial configuration, without prior knowledge of their arrangement.
\item[Assigning particles to nearest centroid:] After initialization, the algorithm computes the distance from each particle to all centroids and assigns each particle to the nearest centroid. This assignment process effectively groups particles based on spatial and momentum proximity, reflecting the notion that particles situated close together in phase space are likely to belong to the same physical region or substructure of the event.
\item[Updating centroids:] Following the assignment of particles to their respective clusters, the centroid of each cluster is recalculated as the average position of all particles contained within that cluster. The contributions of individual particles to the centroid are weighted according to their energies, ensuring that more energetic particles have a greater influence on the centroid's new position.
\item[Iteration:] The aforementioned process of assigning particles to clusters and updating centroids is iterated for a predetermined number of iterations (denoted as \texttt{max\_iters}) or until convergence is achieved, signified by negligible changes in the centroids' positions. This iterative refinement can be interpreted as a progressive enhancement of the initial guesses for the cluster positions, continuing until a stable configuration is reached.
\end{description}

Upon completion of the box or dynamical clustering algorithm, the original particles from each event are grouped into clusters, resulting in a number significantly smaller than the original particle count. This reduction in particle count significantly improves computational efficiency while preserving the essential physics of the system.\footnote{
In principle, after applying clustering within each event, we could extend the method across multiple events to form superclusters. This would further reduce computational complexity by combining representative particles from different events, thereby accelerating the final smearing calculations. However, whether such across-event clustering should be applied depends on whether the physics of the quantities of interest can be preserved.
}

\alglanguage{pseudocode}
\renewcommand{\algorithmicrequire}{\textbf{Input:}}
\renewcommand{\algorithmicensure}{\textbf{Output:}}
\begin{algorithm}[t!]
\small
\caption{$k$-means clustering with energy-weighted centroids}
\label{Algorithm:SimpleKMeansClustering}
\begin{algorithmic}[1]
\Require \texttt{points} - 3D points
\Require \hspace{.75cm} \texttt{energies} - corresponding energy values
\Require \hspace{.75cm} \texttt{k} - number of clusters
\Require \hspace{.75cm} \texttt{max\_iters} - maximum iterations
\Ensure \texttt{clusters} - clusters with updated centroids
\Function{$\mathbf{KMeans}$}{\texttt{points, energies, k, max\_iters}}
    \State \texttt{clusters} $\leftarrow$ initialize \texttt{k} random centroids
    \For {\texttt{each iter} in \texttt{max\_iters}}
        \State \texttt{clear clusters} \Comment{Prepare for new assignments}
        \For {\texttt{each point in points}}
            \State \texttt{assign point to nearest centroid in clusters}
        \EndFor
        \For {\texttt{each cluster}}
            \State \texttt{update centroid as energy-weighted average of points}
        \EndFor
    \EndFor
    \State \Return \texttt{clusters}
\EndFunction
\Statex
\end{algorithmic}
  \vspace{-0.4cm}%
\end{algorithm}

\subsubsection{Representative particles}\label{sec:repparticles}

After the particles are grouped into clusters, a single representative particle is defined for each cluster, and this representative particle is added to the list for tensor calculations. The representative particle of a cluster should preserve the essential physics of the original group. For a cluster of $N$ particles, where the four-momentum of the $i$-th particle is $p^\mu_i = (p^0_i,\, \mathbf{p}_i)$, the four-momentum of the representative particle is defined as:
\begin{equation}\label{eq:psum}
    p^\mu_{\text{rep}} = ( p^0_{\text{rep}},\, \mathbf{p}_{\text{rep}} ) = \sum_{i=1}^{N} p^\mu_i\,,
\end{equation}
where $p^0_{\text{rep}} = \sum_{i=1}^{N} p^0_i$ is the total energy of the system and $\mathbf{p}_{\text{rep}} = \sum_{i=1}^{N} \mathbf{p}_i$ is the total three-momentum of the system. The effective mass $M_{\text{rep}}$ for the representative particle can be calculated from the total four-momentum as $M_{\text{rep}} = \sqrt{(p^0_{\text{rep}})^2 - \mathbf{p}_{\text{rep}}^2}$, which represents the invariant mass of the system. The four-momentum of the representative particle is thus relativistically consistent and behaves correctly under Lorentz transformations.

The four-coordinates of the representative particle can be defined as an energy-weighted average of the four-coordinates of the individual particles within the cluster. This approach is analogous to determining the center of mass in non-relativistic mechanics, where energy serves a role similar to mass:\footnote{
The Newtonian definition of the center of mass can be generalized to the relativistic case in multiple ways, making it a more complex concept \cite{Pryce:1948pf}.
}
\begin{equation}
    x^\mu_{\text{rep}} = (t_{\text{rep}},\, \mathbf{r}_{\text{rep}}) = \frac{\sum_{i=1}^{N} p^0_i x^\mu_i}{\sum_{i=1}^{N} p^0_i}\,,
\end{equation}
where $p^0_i$ is the energy of the $i$-th particle, and $x^\mu_i=(t_i, \mathbf{r}_i)$ denotes the four-coordinates of the $i$-th particle. The summation runs over all $N$ particles in the system, ensuring that particles with higher energy contribute more significantly to the representative position.

\subsection{Tensor calculation}\label{sec:tensorcalc}

In \tensorcalc{}, the calculation of the energy-momentum tensor from discrete particles (original or representative) involves three primary stages, each governed by specific high-level parameters, enabling the investigation of flexible physics scenarios. First, the workflow determines whether to merge multiple particle lists into a single list, treating all particles as if they originated from a single event. This approach simplifies the handling of multiple events but assumes a collective particle system. If merging is not chosen, the workflow processes events individually, with multithreading employed for efficiency. In the second stage, particle clustering is considered to reduce computational cost by grouping particles close in phase space into representative particles. If clustering is enabled, the workflow selects between the two methods, box clustering (regular or dynamical) or dynamical clustering, to calculate representative particles, as discussed in Sec.~\ref{sec:clustering}. If clustering is disabled, the original particles are used without modification. 

\begin{figure}
    \centering
    \includegraphics[width=\linewidth]{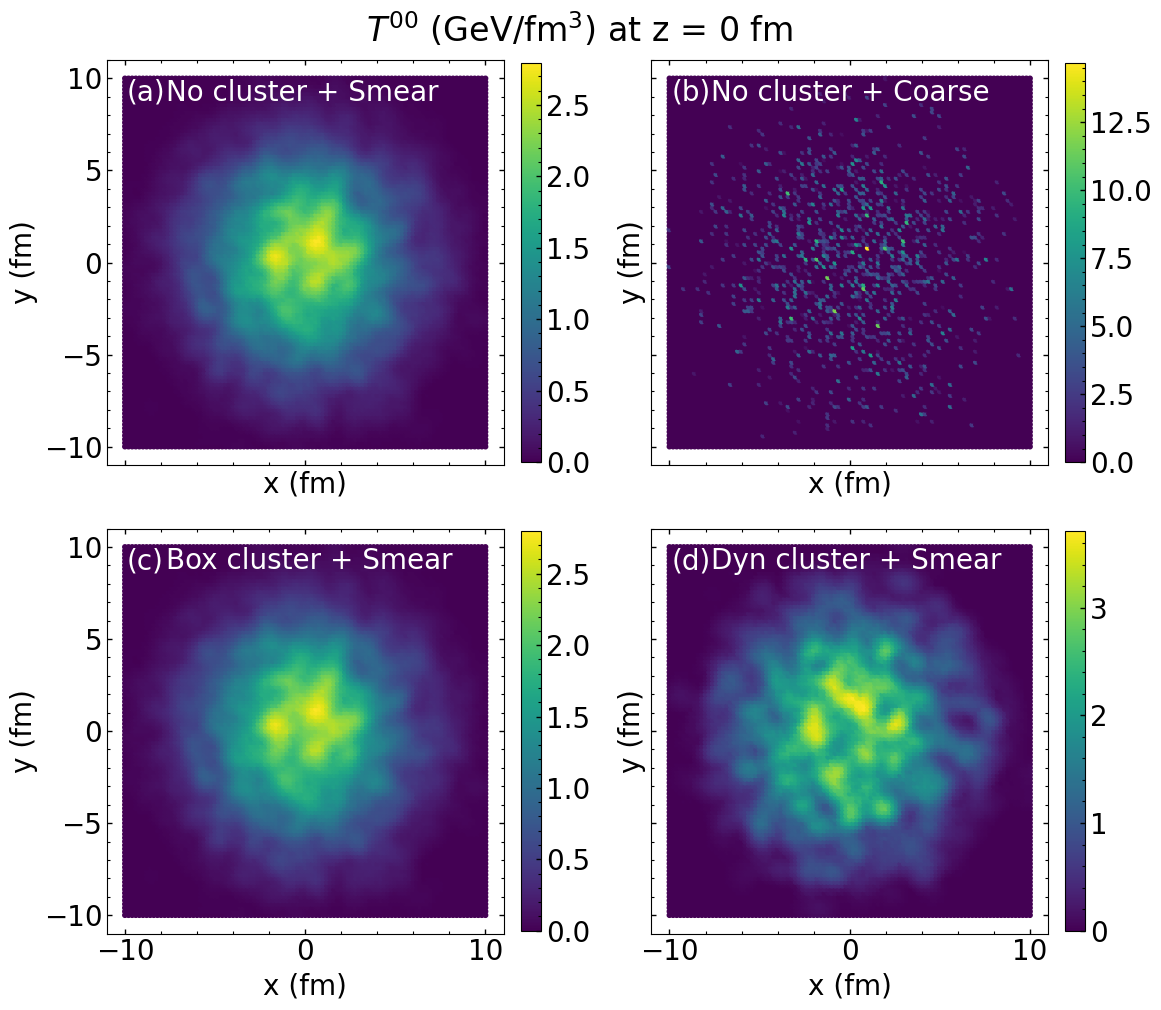}
    \caption{Comparison of $T^{00}$ at $z=0$ fm for different evaluation cases: (a) ``No cluster + Smear'', (b) ``No cluster + Coarse'', (c) ``Box cluster + Smear'', and (d) ``Dyn cluster + Smear''. The case ``Box cluster (reg) + No smear'' (not shown) is visually indistinguishable from (b), while the results for ``Box cluster (reg) + Smear'' and ``Box cluster (dyn) + Smear'' are visually identical to (c). See the text of Sec.~\ref{sec:cases} for detailed case information.}
    \label{fig:compare_tensors}
\end{figure}

Finally, the energy-momentum tensor $T^{\mu\nu}$ and baryon current $J^{\mu}$ are calculated using either coarse-graining, which divides the space-time domain into discrete cells \cite{Endres:2015fna,Oliinychenko:2014tqa,Inghirami:2019muf}, or smearing kernels, which distribute particle contributions smoothly over the domain \cite{Oliinychenko:2015lva,Inghirami:2022afu}. At space-time coordinates $(t, \mathbf{r})$, the energy-momentum tensor is given by:
\begin{equation}\label{eq:Tmunu}
    T^{\mu\nu}(t, \mathbf{r}) = \sum_i \frac{p_i^\mu p_i^\nu}{p_i^0} \, K(t, \mathbf{r}; t_i, \mathbf{r}_i)\;,
\end{equation}
and the baryon current is given by:
\begin{equation}\label{eq:Jmu}
    J^{\mu}(t, \mathbf{r}) = \sum_i b_i\frac{p_i^\mu}{p_i^0} \, K(t, \mathbf{r}; t_i, \mathbf{r}_i)\;.
\end{equation}
Here, the summation is taken over either the original or representative particles, depending on whether clustering is applied; each original or representative particle has space-time coordinates $(t_i, \mathbf{r}_i)$, four-momentum $p_i^\mu = (p_i^0, \mathbf{p}_i)$ and baryon charge $b_i$. The kernel $K(t, \mathbf{r}; t_i, \mathbf{r}_i)$ determines the contribution of a particle at $(t_i, \mathbf{r}_i)$ to the tensor and current at $(t, \mathbf{r})$. It can take the form of a Gaussian-like function for smearing or a delta function for coarse-graining. The coarse-graining method involves dividing the space-time domain into discrete cells and summing the contributions of particles within each cell. For example, the tensor $T^{\mu\nu}$ at a grid point is computed by aggregating the energy-momentum contributions $p_i^\mu p_i^\nu / p_i^0$ from all particles in the corresponding cell, effectively treating the cell as a single unit. Using Eqs.~\eqref{eq:Tmunu} and \eqref{eq:Jmu}, the energy-momentum tensor and baryon current are calculated and presented in Fig.~\ref{fig:compare_tensors} for the various evaluation cases, which are described in detail in Sec.~\ref{sec:cases}.

\begin{figure*}
    \centering
    \includegraphics[width=0.8\linewidth]{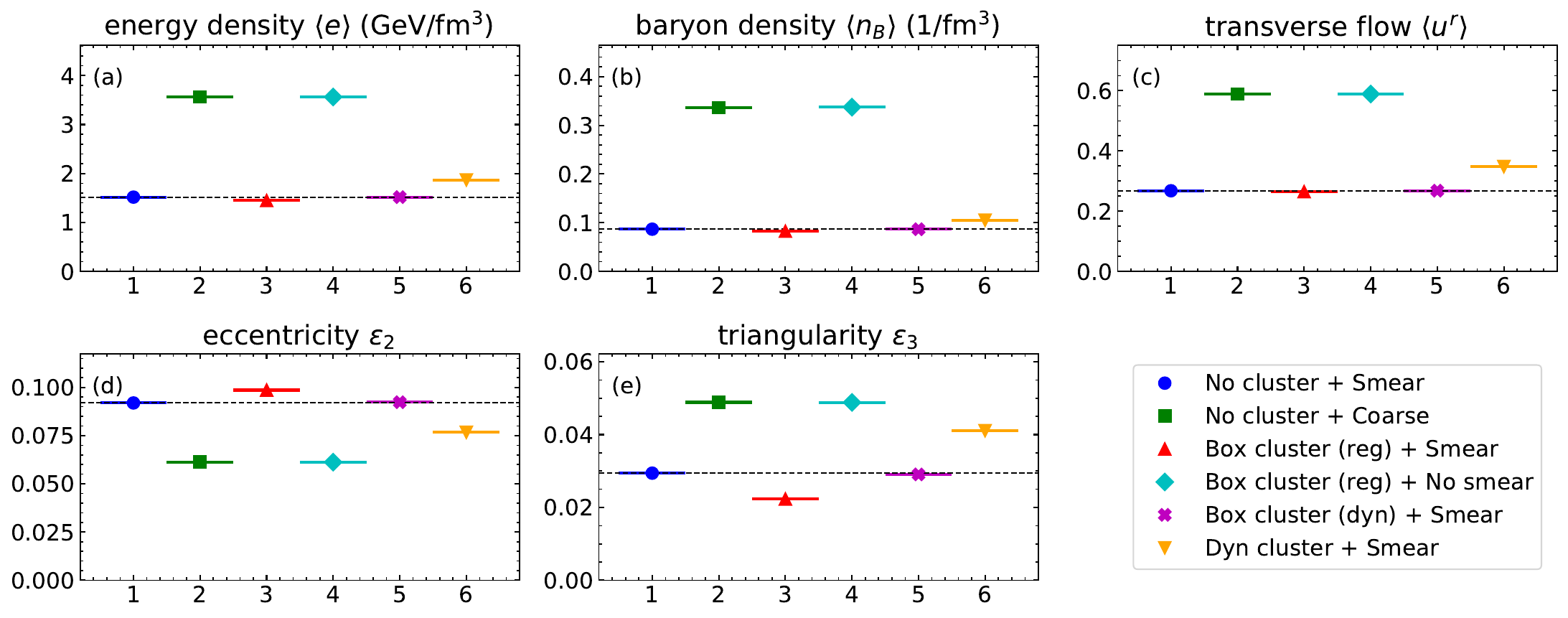}
    \caption{Comparison of key physical quantities at $z=0$ fm for the six evaluation cases: (a) energy density, (b) baryon density, (c) transverse flow, (d) eccentricity, and (e) triangularity. The six cases are represented by distinct markers and colors: ``No cluster + Smear'' (blue circles), ``No cluster + Coarse'' (green squares), ``Box cluster (reg) + Smear'' (red triangles), ``Box cluster (reg) + No smear'' (cyan diamonds), ``Box cluster (dyn) + Smear'' (magenta X's), and ``Dyn cluster + Smear'' (orange triangles pointing down). The dashed line in each plot represents the value corresponding to the baseline case, ``No cluster + Smear.'' See the text for detailed case information.}
    \label{fig:compare_average}
\end{figure*}

\section{Results and discussion}
This section evaluates the physics accuracy of the methods discussed in Sec.~\ref{sec:algorithm} for tensor calculations, with a particular focus on the particle clustering method and the concept of representative particles. Additionally, we discuss the computational efficiency of these methods. All discussions in this section assume a uniform space-time grid for the calculation of the tensor and current.

\subsection{Physics accuracy}\label{sec:physics_accuracy}
\subsubsection{Various evaluation cases}\label{sec:cases}

To evaluate the physics accuracy of the energy-momentum tensor calculation, we compare six distinct cases, each representing a combination of clustering and tensor calculation methods:

\begin{enumerate}[itemsep=0.1em]
\item ``No cluster + Smear'': The original particles are used without clustering, and the tensor is calculated using smearing kernels. This case is physics-motivated and represents the most widely employed approach \cite{Oliinychenko:2015lva,Inghirami:2022afu}. We use it as the baseline for comparison.
\item ``No cluster + Coarse'': The original particles are used without clustering, and the tensor is calculated using coarse-graining \cite{Endres:2015fna,Oliinychenko:2014tqa,Inghirami:2019muf}. It is also a widely used approach for calculating the energy-momentum tensor, particularly when not serving as initial conditions for hydrodynamic simulations.
\item ``Box cluster (reg) + Smear'': Particles are clustered into space-time boxes using the predefined regular grid, and the tensor is calculated using smearing kernels.
\item ``Box cluster (reg) + No smear'': Particles are clustered into space-time boxes using the predefined grid, and the tensor is calculated using coarse-graining.
\item ``Box cluster (dyn) + Smear'': Particles are clustered into space-time boxes using a dynamically generated grid, and the tensor is calculated using smearing kernels.
\item ``Dyn cluster + Smear'': Particles are clustered using dynamical clustering (energy-weighted centroids), and the tensor is calculated using smearing kernels.
\end{enumerate}

These cases allow us to systematically compare the impact of clustering methods and tensor calculation techniques on the accuracy of the energy-momentum tensor and baryon current. For this purpose, we first calculate the tensor and current using Eqs.~\eqref{eq:Tmunu} and \eqref{eq:Jmu} on the predefined regular space-time grid (see Fig.~\ref{fig:compare_tensors}) and then apply the Landau matching condition \cite{landau2013fluid} to obtain energy density, baryon density, and flow velocities. To facilitate a quantitative yet straightforward comparison, we further compute key quantities, such as averaged thermodynamic properties and geometrical anisotropies, as defined in \ref{app:average}.

To ensure meaningful comparisons across these cases, we run SMASH at a fixed center-of-mass energy within a specific impact parameter range and compute all results using consistent grid resolution, size, and smearing parameters where applicable. While we omit exhaustive details here, the key point is to maintain uniformity in relevant factors, enabling a fair and systematic comparison across different evaluation methods. The averaged energy density, baryon density, transverse flow, eccentricity, and triangularity at $z=0$ fm are computed and shown in Fig.~\ref{fig:compare_average}.

\subsubsection{Comparison of tensor calculation methods}

To understand the impact of the two tensor calculation methods---smearing and coarse-graining---we compare cases that use the same set of particles (either original or representative) but differ in how the tensor and current are computed. Specifically, we make two comparisons: No cluster + Smear (case 1 in Fig.~\ref{fig:compare_average}) vs. No cluster + Coarse (case 2 in Fig.~\ref{fig:compare_average}), and Box cluster (reg) + Smear (case 3 in Fig.~\ref{fig:compare_average}) vs. Box cluster (reg) + No smear (case 4 in Fig.~\ref{fig:compare_average}). These comparisons highlight the differences between smearing kernels and coarse-graining when applied to the same set of particles or representative particles.

The results in Fig.~\ref{fig:compare_average} reveal that the cases without smearing exhibit significantly larger average energy and baryon densities, along with higher transverse flow (and much larger standard deviations which are not shown), compared to their smeared counterparts. This is attributed to the preservation of sharper spatial gradients and localized particle contributions in the absence of smearing [see Fig.~\ref{fig:compare_tensors}(b)], resulting in higher peak densities and enhanced anisotropic flow. Interestingly, we observe that smearing enhances eccentricity (second harmonic deformation) while reducing triangularity (third harmonic deformation). The enhancement of eccentricity arises because smearing reinforces the underlying global elliptic shape by distributing contributions more evenly along the large-scale anisotropy. In contrast, triangularity decreases with smearing due to the suppression of localized ``hot spots'' or triangular perturbations, which are critical for generating small-scale deformations \cite{Qiu:2011hf}.

These findings highlight the competing effects of smoothing techniques in modifying the spatial and dynamic characteristics of the system, offering insights into the impact of smearing and coarse-graining on observables in heavy-ion collision simulations. For example, this implies that for electromagnetic emission calculations with the bulk evolution simulated by transport models \cite{Endres:2015fna,Savchuk:2022aev}, the results could be strongly affected by whether smearing is applied to the underlying particle degrees of freedom when obtaining thermodynamic properties. Similarly, investigating the freeze-out line defined in terms of local energy density or temperature depends on the choice of tensor calculation method  \cite{Oliinychenko:2014tqa,Inghirami:2019muf}.

\subsubsection{Effectiveness of representative particles}

In Sec.~\ref{sec:repparticles}, we introduced the concept of representative particles, which are used to replace groups of particles that fall into the same cluster to reduce the computational cost of tensor calculations. To evaluate the effectiveness of representative particles in preserving the underlying physics, we compare the ``No cluster + Coarse'' and ``Box cluster (reg) + No smear'' cases. These cases are conceptually similar, as both involve grouping particles into space-time boxes defined by a regular grid and calculating the energy-momentum tensor without smearing. However, they differ in whether representative particles are used.

In ``No cluster + Coarse'', the tensor $T^{\mu\nu}$ is calculated by summing the contributions $T^{\mu\nu} = p^\mu p^\nu / p^0$ from all original particles within each box. This approach directly aggregates the tensor contributions of individual particles. However, in ``Box cluster (reg) + No smear'', a representative particle is first created for each box by summing the four momenta $p^\mu$ of all particles within the box [see Eq.~\eqref{eq:psum}]. The tensor $T^{\mu\nu}$ at a space-time point is then calculated for the single representative particle located at that point as $T^{\mu\nu} = p_{\text{rep}}^\mu p_{\text{rep}}^\nu / p_{\text{rep}}^0$. This approach effectively replaces the ensemble of particles within a box with a single representative particle.

While these two cases should yield very similar results, they are not exactly equivalent due to the nonlinearity of the tensor calculation. Specifically, the tensor $T^{\mu\nu} = p^\mu p^\nu / p^0$ is not linear in $p^\mu$, meaning that summing $T^{\mu\nu}$ over particles is not identical to calculating $T^{\mu\nu}$ for the summed momenta and energy. Comparing these two cases provides a valuable test of the idea of representative particles. Indeed, their tensors are visually indistinguishable [see Fig.~\ref{fig:compare_tensors}(b)], and the averaged thermodynamic properties and geometrical anisotropies are nearly identical, as demonstrated by cases 2 and 4 in Fig.~\ref{fig:compare_average}. This comparison confirms the ability of representative particles to accurately capture the physics of the original particle ensemble. This result is consistent with the physical motivation for using representative particles, as discussed in Sec.~\ref{sec:clustering}.

\subsubsection{Comparison of clustering methods}

We assess the impact of different clustering methods on preserving the physics accuracy by comparing the following cases to the baseline (``No cluster + Smear''): ``Box cluster (reg) + Smear'', ``Box cluster (dyn) + Smear'' (case 5 in Fig.~\ref{fig:compare_average}), and ``Dyn cluster + Smear'' (case 6 in Fig.~\ref{fig:compare_average}). All cases use the same smearing method to calculate the energy-momentum tensor and baryon current from either the original or representative particles.

As shown in Fig.~\ref{fig:compare_average}, all three clustering methods with smearing yield results closer to the baseline compared to the cases without smearing. However, the two box clustering methods---using either a regular grid or a dynamically generated grid---perform better than the dynamical clustering case introduced in Sec.~\ref{sec:dynclustering}. Specifically, the dynamical clustering case results in a system with higher energy and baryon densities, as well as stronger transverse flow at $z = 0$. This behavior can be attributed to the higher densities at $z = 0$, which cause particles away from $z = 0$ to be attracted toward centroids near $z = 0$ during the clustering process, further increasing the densities in this region.

In contrast to the dynamical clustering method, the two box clustering methods group particles in a more controlled manner, constrained by the space-time grid boxes. Notably, the box clustering method with an adaptive grid produces results almost identical to the baseline. This is expected, as this method aligns more naturally with the actual particle density, as discussed in Sec.~\ref{sec:boxclustering}. Therefore, from a physics perspective, box clustering with an adaptive grid followed by smearing is the most accurate among the evaluated cases when compared to the baseline. In the next section, we will demonstrate that this method is also computationally efficient.

\subsection{Computational efficiency}\label{sec:computational_efficiency}

\subsubsection{Computational cost for tensor calculation}

The computational cost of calculating the energy-momentum tensor and baryon current depends on two major steps: (1) clustering particles (if applicable) and (2) computing contributions from particles or representative particles to the tensor and current at each space-time grid point. For cases involving smearing, the second step involves looping over grid points and particles. For cases involving coarse-graining, the second step typically involves checking which particles fall into each grid box and summing their contributions.

The clustering step, when applied, reduces the number of particles from $N_{\text{particles}}$ to $K_{\text{rep}}$ (where $K_{\text{rep}} \ll N_{\text{particles}}$), significantly lowering the computational cost of the smearing step. However, the efficiency of clustering varies depending on the method used (see Sec.~\ref{sec:clustering}). In the tensor calculation step, the computational cost is dominated by the number of particle-grid pairs. For smearing, this scales as $O(K_{\text{rep}} \times M_{\text{grid}})$ when using representative particles, compared to $O(N_{\text{particles}} \times M_{\text{grid}})$ for the naive approach. For coarse-graining, the cost depends on the number of particles per grid box, which can vary significantly depending on the clustering method and particle distribution. 

Overall, the choice of clustering and tensor calculation methods introduces trade-offs between physics accuracy and computational efficiency. In Sec.~\ref{sec:physics_accuracy}, we explored the physics accuracy, and next we compare the computational costs of the key cases, focusing on their efficiency and scalability.

\subsubsection{Comparison of computational costs}

The two conventional methods, ``No cluster + Smear'' and ``No cluster + Coarse'', serve as benchmarks for computational cost. ``No cluster + Smear'' involves evaluating contributions from all original particles within a certain distance of each grid point, leading to a complexity of $O(N_{\text{particles}} \times M_{\text{grid}})$. In contrast, ``No cluster + Coarse'' sums contributions from particles within each grid box, which involves looping over all original particles for each grid point, resulting in a computational complexity of $O(N_{\text{particles}} \times M_{\text{grid}})$. While this approach is not computationally more expensive than smearing, it results in less accurate physics, as shown in Sec.~\ref{sec:physics_accuracy}.

The three cases, ``Box cluster (reg) + Smear'', ``Box cluster (dyn) + Smear'' and ``Dyn cluster + Smear'', apply smearing using representative particles, but their clustering methods differ significantly. In the ``Box cluster (reg) + Smear'' case, a predefined grid is used for both clustering and tensor calculation. This results in a clustering complexity of $O(N_{\text{particles}} \times M_{\text{grid}})$, as particles are assigned to fixed grid cells, and a smearing complexity of $O(M_{\text{grid}} \times M_{\text{grid}})$, since the number of representative particles $K_{\text{rep}}$ matches the number of grid points $M_{\text{grid}}$. In contrast, the ``Box cluster (dyn) + Smear'' case dynamically clusters particles using hash maps, reducing the clustering complexity to $O(N_{\text{particles}})$. When the smearing step is performed on the predefined grid, its complexity is $O(K_{\text{rep}} \times M_{\text{grid}})$, where $K_{\text{rep}} \ll N_{\text{particles}}$ in most cases. While ``Dyn cluster + Smear'' adapts to the particle distribution using $k$-means clustering to reduce the particle count, it introduces additional overhead due to the iterative nature of the algorithm, scaling as $O(N_{\text{particles}} \log N_{\text{particles}})$. This makes it less efficient than ``Box cluster (dyn) + Smear'', especially for large systems.

\begin{figure}
    \centering
    \includegraphics[width=\linewidth]{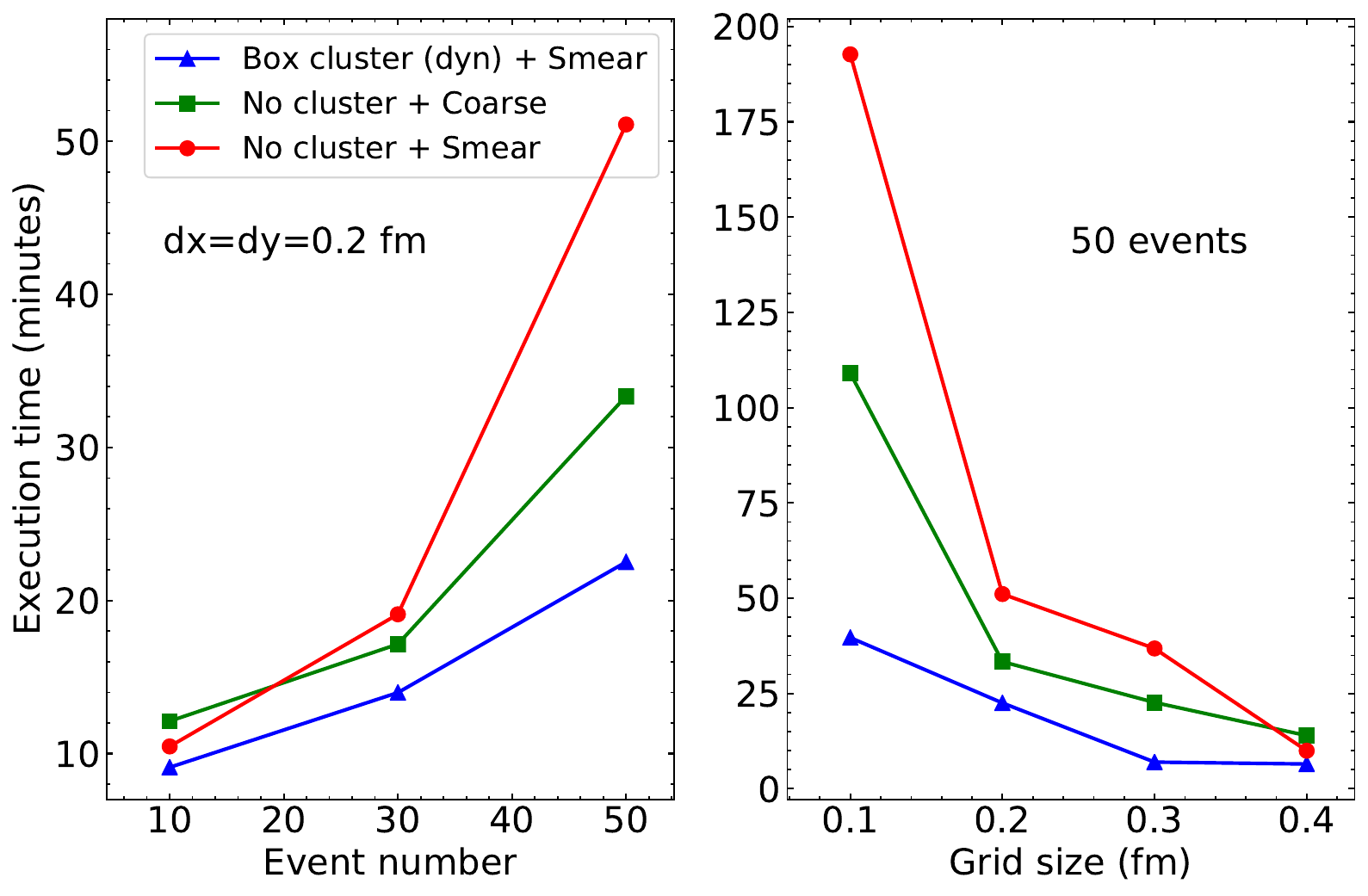}
    \caption{Execution time scalability comparison for different tensor computation methods. The left panel shows execution time in minutes as a function of the number of events, with a fixed grid size of $\text{d}x=\text{d}y=0.2$ fm. The right panel presents execution time as a function of grid size ($\text{d}x, \text{d}y$) while keeping the number of events fixed at 50. All calculations are performed using 8 threads on a computing cluster.}
    \label{fig:scalability}
\end{figure}

The ``Box cluster (dyn) + Smear'' approach benefits significantly from the reduction in particle count during clustering, particularly for large-scale heavy-ion simulations where $N_{\text{particles}}$ can be on the order of millions. The computational savings achieved by ``Box cluster (dyn) + Smear'' are substantial, making it a more efficient choice for such systems. 
The memory usage of ``Box cluster (dyn) + Smear'' is also optimized, as it only stores non-empty grid cells, significantly reducing memory overhead compared to ``Box cluster (reg) + Smear''. 
Furthermore, using hash maps in the clustering step ensures that particle grouping and representative particle creation are efficient, with average $O(1)$ complexity for insertions and lookups per particle. This makes the clustering method highly scalable and suitable for high-performance computing environments. Benchmarks with increasing event counts and decreasing grid sizes, shown in Fig.~\ref{fig:scalability}, demonstrate that ``Box cluster (dyn) + Smear'' scales more efficiently than the two conventional methods, particularly for large systems.

\subsection{Future improvements}
As illustrated by the previous two subsections, among the evaluated cases, ``Box cluster (dyn) + Smear'' stands out as both computationally efficient and physics-accurate. It achieves perfect agreement with the baseline (``No cluster + Smear'') while significantly reducing computational costs through particle clustering and efficient hash map usage. This makes it the preferred method for large-scale simulations, where balancing accuracy and efficiency is critical.

While the current implementation of ``Box cluster (dyn) + Smear'' uses a predefined grid for smearing, further improvements in computational efficiency can be achieved by transitioning to an adaptive grid that leverages information from the clustered particles.\footnote{
In contrast to the case described in Sec.~\ref{sec:adaptive}, where grid boundaries are adjusted but local resolution remains fixed at each time step, the adaptive grid discussed here dynamically adjusts local resolution based on particle densities.
}
An adaptive grid would dynamically adjust its resolution based on the density of representative particles, focusing computational resources on regions of interest where particle density is high. This approach would reduce the number of grid points $M_{\text{grid}}$ to $M'_{\text{grid}}$, where $M'_{\text{grid}} \ll M_{\text{grid}}$ in sparse regions, further lowering the computational complexity of the smearing step to $O(K_{\text{rep}} \times M'_{\text{grid}})$. The adaptive grid could be constructed using the spatial distribution of representative particles, ensuring that high-density regions are finely resolved while low-density regions are coarsely sampled. This would not only improve computational efficiency but also enhance the accuracy of the energy-momentum tensor calculation in regions where it matters most. Implementing such an adaptive grid would require additional algorithmic development, but the potential gains in efficiency and accuracy make it a promising direction for future work.

\section{Summary}

In this study, we introduced the MATRICS framework, designed to efficiently calculate the space-time distribution of the energy-momentum tensor, charge currents, and thermodynamic properties from discrete particles generated by transport models in heavy-ion collision simulations. By leveraging advanced techniques such as adaptive grids, particle clustering, and parallel processing, we addressed the computational challenges of obtaining continuous space-time distributions of thermodynamic properties from transport models. The framework's modular design and innovative algorithms enable high-precision calculations of thermodynamic and collective behavior while significantly reducing computational costs.

A key feature of MATRICS is its implementation of particle clustering methods to reduce particle counts by grouping particles into representative particles. This coarse-graining technique, combined with smearing kernels, maintains physical accuracy while optimizing computational efficiency. Our results demonstrate that the ``Box cluster (dyn) + Smear'' method closely aligns with the physics-motivated baseline approach (``No cluster + Smear'') while achieving substantial computational savings, making it ideal for large-scale simulations aimed at understanding the event-averaged bulk properties of the baryon-rich QCD matter created in heavy-ion collisions.

Future improvements to MATRICS could focus on further optimizing the adaptive grid by dynamically adjusting its resolution based on the distribution of representative particles. This would enhance computational efficiency and accuracy, particularly in high-density regions, and is especially useful when the goal is not to provide initial conditions for hydrodynamic evolution, which typically require a regular grid. Additionally, exploring across-event clustering and further parallelization could improve the framework's scalability and applicability to even larger-scale simulations.

In conclusion, the MATRICS framework represents an advancement in extracting thermodynamic properties from transport models of heavy-ion collisions. It provides a powerful tool for exploring the properties of QCD matter under extreme conditions, particularly in high baryon density regions. Its flexibility, efficiency, and accuracy make it a valuable resource for the nuclear physics community, with potential applications extending to other areas of high-energy physics and beyond.

\section*{Acknowledgements}
The author \cite{fnackn} acknowledges helpful conversations with Tom Reichert and Shuzhe Shi. The author also extends gratitude to Shuzhe Shi for providing insightful comments on the manuscript. Computations were made on the computers managed by the Ohio Supercomputer Center \cite{OhioSupercomputerCenter1987}.


\appendix

\section{Averaged quantities and anisotropies}\label{app:average}

To evaluate the results of calculating the energy-momentum tensor and baryon current using different particle clustering and tensor calculation methods, we investigate several averaged thermodynamic quantities and geometrical anisotropies \cite{Shen:2013vja,Churchill:2023zkk,Churchill:2023vpt}. These quantities are computed using weighted sums over the transverse plane at a specific time step for a fixed $z$, accounting for the spatial distribution of energy density and flow velocity. The weights are based on the local energy density $ e $ and the time component of the flow velocity $ u^0 $ at each spatial grid point.

To calculate these quantities, we first apply Landau matching to obtain local thermodynamic quantities and flow four velocities from the energy-momentum tensor and baryon current. Then the average energy density $ \langle e \rangle $ is computed as $\langle e \rangle = \sum_i w_i e_i/\sum_i w_i$, where the weight $ w_i $ for each grid point is given by $w_i = e_i u^0_i$. Here $ e_i $ is energy density at grid point $ i $, $ u^0_i $ the time component of the flow velocity at grid point $ i $ (Lorentz gamma factor). The summation is performed over all the fluid cells in the transverse plane. Other averaged quantities, such as baryon density and transverse flow, are calculated similarly using the same weighting scheme. By incorporating the weights $ w_i $, the contributions from regions with higher energy density are emphasized \cite{Shen:2013vja,Du:2023gnv,Du:2023efk}.

For each harmonic $ n $, the geometrical anisotropy $ \epsilon_n $ is defined as:
\begin{equation}
    \epsilon_n = \frac{\sqrt{\left( \sum_i w_i r_i^n \cos(n\phi_i) \right)^2 + \left( \sum_i w_i r_i^n \sin(n\phi_i) \right)^2}}{\sum_i w_i r_i^n}\,,
\end{equation}
where $ r_i = \sqrt{x_i^2 + y_i^2} $ is the transverse radius at grid point $ i $,
$ \phi_i$ the azimuthal angle at grid point $ i $. The numerator quantifies the anisotropic deformation in harmonic $ n $, while the denominator normalizes the anisotropies to account for the total weight. We note that a more rigorous definition of the anisotropies would require calculating the transverse radius with respect to the center of mass and aligning event-plane angles of all the events before averaging. However, for the purpose of comparing the energy-momentum tensor and baryon current obtained from different methods, the current definition suffices.

\bibliographystyle{elsarticle-num} 
\bibliography{refs}

\begin{thebibliography}{10}
\bibitem{fnackn}
Although the journal’s policy does not require disclosure of grammar and
  spelling correction tools, we note that AI-assisted tools (e.g., Grammarly
  and ChatGPT) were used to improve the clarity and readability of the
  manuscript. All revisions were carefully reviewed to ensure that the intended
  meaning and scientific content remained unchanged.
\end{thebibliography}

\end{document}